\documentclass[11pt]{article}
\usepackage{booktabs}
\usepackage{multirow,hyperref}
\usepackage{footnote}
\usepackage[normalem]{ulem}
\RequirePackage[OT1]{fontenc}
\usepackage{times,graphicx,theorem,epsfig,lscape,amsmath,dcolumn,multirow,epic,float,enumerate,comment,amssymb}
\usepackage[round]{natbib}
\usepackage{etoolbox}

\makeatletter

\pretocmd{\NAT@citex}{%
	\let\NAT@hyper@\NAT@hyper@citex
	\def\NAT@postnote{#2}%
	\setcounter{NAT@total@cites}{0}%
	\setcounter{NAT@count@cites}{0}%
	\forcsvlist{\stepcounter{NAT@total@cites}\@gobble}{#3}}{}{}
\newcounter{NAT@total@cites}
\newcounter{NAT@count@cites}
\def\NAT@postnote{}

\def\NAT@hyper@citex#1{%
	\stepcounter{NAT@count@cites}%
	\hyper@natlinkstart{\@citeb\@extra@b@citeb}#1%
	\ifnumequal{\value{NAT@count@cites}}{\value{NAT@total@cites}}
	{\ifNAT@swa\else\if*\NAT@postnote*\else%
		\NAT@cmt\NAT@postnote\global\def\NAT@postnote{}\fi\fi}{}%
	\ifNAT@swa\else\if\relax\NAT@date\relax
	\else\NAT@@close\global\let\NAT@nm\@empty\fi\fi
	\hyper@natlinkend}
\renewcommand\hyper@natlinkbreak[2]{#1}

\patchcmd{\NAT@citex}
{\ifNAT@swa\else\if*#2*\else\NAT@cmt#2\fi
	\if\relax\NAT@date\relax\else\NAT@@close\fi\fi}{}{}{}
\patchcmd{\NAT@citex}
{\if\relax\NAT@date\relax\NAT@def@citea\else\NAT@def@citea@close\fi}
{\if\relax\NAT@date\relax\NAT@def@citea\else\NAT@def@citea@space\fi}{}{}

\makeatother

\usepackage{threeparttable}
\usepackage{rotating}
\usepackage{lipsum}
\usepackage{comment,bm}
\usepackage{afterpage}
\RequirePackage{url}
\usepackage{placeins}
\usepackage{color}
\newtheorem{assumption}{Assumption}

\newcommand{\TT}{\mbox{\tiny{ATT}}}
\newcommand{\MTE}{\mbox{\tiny{MTE}}}
\newcommand{\PRTE}{\mbox{\tiny{PRTE}}}

\newcommand{\ind}{\perp \! \! \! \perp}

\newcommand{\blinding}[2]{#1}   

\begin{document}

\begin{center}
	
	{\Large Is being an only child harmful to psychological health?: Evidence 
	from an instrumental variable analysis of China's One-Child Policy\\}
	\medskip
	\blinding{
		Shuxi Zeng \quad Fan Li \quad Peng Ding
		\footnote{Shuxi Zeng is Ph.D. student, Department of Statistical Science, Duke University, Durham, NC, 27705 (email: shuxi.zeng@duke.edu); Fan Li is Associate Professor, Department of Statistical Science, Duke University, Durham, NC, 27705 (email:
			fl35@duke.edu); Peng Ding is Assistant Professor, Department of Statistics, University of California, Berkeley, CA 94720 (email: pengdingpku@berkeley.edu).}
		
	}
	\date{}
	
\end{center}
%
%
%
%
{\centerline{ABSTRACT}
This paper evaluates the effects of being an only child in a family on 
psychological health, leveraging data on the One-Child Policy in China. We use 
an instrumental variable approach to address the potential unmeasured 
confounding between the fertility decision and psychological health, where the 
instrumental variable is an index on the intensity of the implementation of the 
One-Child Policy. We establish an analytical link between the local 
instrumental variable approach and principal stratification to accommodate the 
continuous instrumental variable. Within the principal stratification 
framework, we postulate a Bayesian hierarchical model to infer various causal 
estimands of policy interest while adjusting for the clustering data structure. 
We apply the method to the data from the China Family Panel Studies and find 
small but statistically significant negative effects of being an only child on 
self-reported psychological health for some subpopulations. Our analysis 
reveals treatment effect heterogeneity with respect to both observed and 
unobserved characteristics. In particular, urban males suffer the most from 
being only children, and the negative effect has larger magnitude if the 
families were more resistant to the One-Child Policy. We also conduct 
sensitivity analysis to assess the key instrumental variable assumption.
	\vspace*{0.3cm}
}
\clearpage

\section{Introduction}

The One-Child Policy (OCP) was a birth planning policy in China to control the 
rapid population growth during the mid 20th century. From late 1979 to 2015, 
the Chinese government enforced strict regulations to limit the number of 
children that each family could have. In most cases, each family was allowed to 
have only one child. A range of penalties were imposed on families who violated 
the OCP, including hefty financial fines, restriction to education, and 
demotion of parents working in the public sector. The OCP formally ended in 
2015 when each family was allowed to have two children. As the world's most 
aggressive policy of population planning, the OCP had far-reaching and 
transformative influence on every facet of the Chinese society. Although the 
OCP was effective in controlling the population growth, its social, economic 
and cultural impact remains controversial. Various aspects of the OCP have been 
studied in the literature. For example, a stream of research focused on its 
demographic influence, such as the artificial selection of the gender of 
newborns or ``man-made'' twins \citep[e.g.][]{Ebenstein2010, Huang2016}, 
marriage distortion, and welfare loss \citep{huang2015one}. Another strand of 
research studied its impact on human capital accumulation by inducing more 
families to have a single child, e.g., whether the OCP leads to higher 
educational attainment or better physical conditions of the children 
\citep[e.g.][]{Rosenzweig2009, Qian2009, Liu2014, huang2016fertility}.

This paper investigates the causal effects of being an only child on subjective 
psychological well-being. This is an important social topic because a prevalent 
yet unsubstantiated perception in China is that the only children are more 
selfish, insecure and immature compared to the children with siblings. Indeed, 
the only children following the OCP in China are often called the generation of 
``little emperors" \citep{Cameron2013}. The effects of being an only child have 
been studied in different disciplines from complementary perspectives. Research 
in psychology has suggested that sibling companion contributes positively to 
the development of psychological health \citep{Dunn1988, Brody2004,Mchale2012}. 
On the other hand, in labor economics,  \cite{becker1973interaction} proposed 
the influential theoretical model of ``quantity-quality interaction" for 
fertility choice, which explains the relationship between the size of a family 
(i.e. ``quantity'') and the well-being (i.e. ``quality'', such as physical 
health and education attainment) of each child. Some empirical studies found a 
negative association between the family size and school achievement 
\citep[e.g.,][]{leibowitz1974home,hanushek1992improving}, but such an 
association was not observed in other quasi-experiments 
\citep[e.g.,][]{black2005more,angrist2010multiple}. \cite{mogstad2016testing} 
revealed a more complex and heterogeneous effect of the number of children on 
children's well-being. These studies found heterogeneous patterns across 
different populations. It is therefore of interest to extend the research to 
the Chinese population given its different cultural background.

Evaluating the effects of being an only child on psychological health is 
challenging because
some unmeasured confounding factors may simultaneously affect the fertility
decision and psychological health outcomes of the children within a family. The 
families with
only one child may differ from the families with more than one child in
systematic but unobserved ways (e.g. nurturing environment). A credible causal
comparison requires proper adjustment of both measured and unmeasured
confounding. A number of studies leveraged the OCP as an exogenous shock to
address confounding. \cite{Cameron2013} adopted an instrumental variable
(IV) approach with the birth year as an IV, leveraging a natural experiment on
individuals who were born just before and just after the
implementation of the OCP. Based on measures of subjective well-being, they 
concluded
that only children are less trustworthy, more pessimistic and risk taking.
Another approach is to use the implementation intensity of the OCP as an IV.
More specifically, although the OCP was a national policy, it was implemented
with different intensities across regions and time periods.
Some researchers adopted the fine
rate  as a proxy for the intensity \citep{Ebenstein2010, Huang2016}.
\cite{Attane2002} developed the Indicator of Family Planning Policy Resistance
(IFPPR)---a continuous positive index---to characterize the OCP implementation
intensity. From a design perspective, the variation in the implementation
intensity provides a natural experiment on fertility choice. Along this
line of thoughts, \cite{Wu2014} used the IFPPR as an IV to analyze a
subsample of the China Family Panel Studies (CFPS) \citep{Xie2014}, and found
that being an only child had a negative effect on self-reported psychological
health. \cite{Wu2014} employed a three-stage-least-squares method
\citep{heckman1978dummy}, assuming a homogeneous treatment effect.
Wu's analysis left three important statistical challenges open. First, the
three-stage-least-squares model lacks a formal causal interpretation when the
strong assumption of homogeneous treatment effect does not hold. Second, the 
sample
units are clustered within provinces; importantly, the IV---the OCP
implementation density---is measured at the cluster  level defined by the 
province of birth. Third,
the key IV assumptions such as exclusion restriction may be violated to a
certain degree and its impact to the analysis remains unknown.

Motivated by these limitations in \citet{Wu2014}, we propose a set of new 
methods for continuous IV analysis. Specifically, under the potential outcomes 
framework to causal inference, we extend the local IV method 
\citep{Heckman1999, Heckman2001policy}, capitalizing on an intrinsic link 
between local IV and principal stratification \citep{Frangakis2002principal}. 
This link was implicitly implied in \cite{heckman2018} but has not been used in 
statistical analysis previously. Based on this link, we translate the causal 
estimands in the local IV method into the causal parameters under principal 
stratification. Within the principal stratification framework, we propose a 
flexible Bayesian model that allows for heterogeneous treatment effects and 
accommodates the clustering structure and ordinal outcomes, as well as provides 
straightforward posterior inference of the causal estimands. Our method extends 
the binary IV approach for randomized experiments with noncompliance 
\citep{Angrist1996}, and the principal stratification approach to clustered 
randomized trials \citep{Frangakis2002clustered, Booil2008, 
forastiere2016identification}. Moreover, we propose a sensitivity analysis 
method to assess the potential impact of the violation of the key exclusion 
restriction assumption, which supplements the existing literature on IV 
sensitivity analysis that has largely focused on linear models with constant 
effects \citep[e.g.][]{small2007sensitivity,wang2018sensitivity}. Although our 
methods are originally motivated by the empirical application of OCP, they are 
applicable to a wide range of studies with continuous IV.

%

Our local IV analysis stratifies on two key background covariates, leading to 
four subpopulations defined by male or female and urban or rural areas. Across 
three self-reported measures of confidence, anxiety and desperation, being an 
only child does not affect the rural populations significantly but does exert 
small yet significant negative effects on the urban subpopulations. Thanks to 
the local IV method, we also detect that urban males suffered the most from 
being only children especially for those from families which were more 
resistant to the OCP. We offer possible explanations to the treatment effect 
heterogeneity and also provide several sensitivity checks and simulation 
evaluations.  The data and programming code of this paper is available at:

\url{https://github.com/zengshx777/OCP_LIV_JRSSA}.

\section{The Causal Inference Framework}\label{sec:methods}

\subsection{Basic setup and assumptions}

Consider a sample of $N$ units from a population of interest, where the $i$th
($i=1,2,\cdots,N$) individual belongs to a cluster $C_{i}$
($C_{i}=1,2,\cdots,G$) defined by the province of birth. For individual $i$,
we observe a set of pretreatment
covariates $X_{i}$, for example, the demographic information and family
background. Let $T_{i}$ be the treatment indicator, with $T_{i}=1$ if
individual $i$ is an only child and $T_{i}=0$ otherwise. We
observe a response variable $Y_{i}$ and a continuous IV $Z_{i}$ bounded between 
$z_{\min}$ and $z_{\max}$. In our application,
$Y_{i}$ is an ordered self-reported measurement of psychological health with a
larger value representing a better condition, and $Z_{i}$ is the IFPPR index,
with a larger value indicating a higher policy implementation intensity (i.e., 
it is decreasing in the original index created by \citet{Attane2002}).

We proceed under the potential outcomes framework. We invoke the Stable Unit 
Treatment Value Assumption (SUTVA) \citep{Rubin1980}. Specific to our 
application, SUTVA implies two components. First, whether individual $i$ is an 
only child depends on the IFPPR in the corresponding province, but does not 
depend on the IFPPR in other provinces. Second, the IFPPR and the fertility 
decision for one family do not affect the psychological health of children from 
other families. Both assumptions are reasonable in our application. SUTVA 
allows us to write $T_{i}(z)$ as the potential treatment status had the IV of 
unit $i$ taken the value $z$, and $Y_{i}(z,t)$ as the potential outcome had 
unit $i$ been exposed to the IV value $z$ and the treatment status $t$. For 
each individual $i$, we observe the treatment status $T_{i}=T_{i}(Z_{i})$ and 
the outcome $Y_{i}=Y_{i}(Z_{i},T_{i}(Z_{i}))$.

We now formally introduce the IV assumptions.
\begin{assumption}
	\label{Unconfoundedness}
	$
	Z_{i} \ind \{Y_{i}(z,t),T_{i}(z) : t=0,1,z\in [z_{\min},z_{\max}]\}\mid
	X_{i}.
	$
\end{assumption}

Assumption \ref{Unconfoundedness} requires that the IV is randomly assigned
with respect to the potential treatment status and outcomes, conditioning on
the covariates $X_{i}$. This assumption is reasonable in our application
because the intensity of implementing OCP largely depended on the specific
province, and which province a family resided in is random conditional on the
family background.

\begin{assumption} \label{as:nontrivial} The probability $\Pr(T_{i}=1 \mid
	Z_{i}=z, X_i )$ is a non-degenerate function of $z$ for all $X_i$.
\end{assumption}

Assumption \ref{as:nontrivial} requires that the IV has non-zero effects on the
treatment assignment, that is, the probability of being an only child varies 
with the intensity of the OCP conditioning on covariates.
This assumption is testable by checking the empirical
distribution of the treatment conditional on the covariates. We will present
such evidence in Section \ref{sec:app}.

\begin{assumption} [Monotonicity]\label{monotonicity}
	For any $z<z'$, $T_{i}(z)\leq T_{i}(z')$ for all $i=1,...,N$.
\end{assumption}

Assumption \ref{monotonicity} requires that the IV monotonically affects the
potential treatment status $T_{i}(z)$; it extends the monotonicity assumption
on a binary IV in \cite{Angrist1996}. In our application, monotonicity assumes
that increasing the intensity of implementing OCP will not increase the number
of children a family had, which is reasonable given the severe financial and
social punishment incurred from violating the OCP.

\begin{assumption} [Exclusion Restriction]\label{ER} The potential outcomes is
	solely a function of $t$, that is, for any unit $i$ and for any $z, z'$,
	$
	Y_{i}(z,t)=Y_{i}(z',t),
	$
	for  $ t=0$ and $1$.
\end{assumption}

Assumption \ref{ER} requires that the IV affects the outcomes only through its
effects on the treatment. In our application, this means that the intensity of
the OCP only affects whether a child is an only child or not, but does not
directly affect psychological health. Namely, given the same treatment status,
different intensity of implementing OCP has no influence on the outcomes.
Under exclusion restriction, we can use the single-index notation for the
potential
outcome, $Y_{i}(t)$, instead of the double-index $Y_{i}(z,t)$, which we adopt 
hereafter.
Exclusion restriction is a crucial identification assumption in the IV analysis
but might be questionable in practice. Intensity of the implementation of the 
OCP might have other channels to affect psychological health. For instance, it 
might change the divorce rate, which further affects psychological health of 
children. This could lead to violation of exclusion restriction, although we 
find the divorce rate was low throughout the study period.
%
In Section \ref{sec:SA}, we will perform
a sensitivity analysis to examine the consequences of potential violation to
exclusion restriction in our application.

\subsection{Causal estimands}	

We now introduce three causal estimands in the context of a continuous IV. The
first estimand is the standard average treatment effect on the treated (ATT):
\begin{eqnarray}
\label{TT_average}
\tau^{\textup{\TT} } =  E\{Y_{i}(1)-Y_{i}(0)\mid T_{i}=1 \}.
\end{eqnarray}

The second estimand is the policy-relevant treatment effect (PRTE)
\citep{Heckman2001policy}:
\begin{eqnarray}
\label{PRTE_average}
\tau^{\textup{\PRTE}} =  E\{Y_{i}(1)-Y_{i}(0)\mid
T_{i}(z_{\min})=0,T_{i}(z_{\max})=1 \} .
\end{eqnarray}
This estimand measures the causal effect for children from families who would
give birth to more than one child at the lowest policy intensity but only one
child at the highest policy intensity. These families changed their fertility 
decisions because of the OCP. The estimand
$\tau^{\textup{\PRTE}} $ quantifies the causal effect for a
subpopulation defined by the joint potential treatment status $T_{i}(z)$, where
$T_{i}(z)$ measures the inclination to receive the treatment.

The third estimand is the marginal treatment effect (MTE):
\begin{eqnarray}
\label{MTE_average}
\tau^{\textup{\MTE}}(z) =  E\{Y_{i}(1)-Y_{i}(0)\mid T_{i}(z)=1 \text{ and
}T_{i}(z')=0 \textup{ for any }z'<z \}.
\end{eqnarray}
It measures the causal effect for the units at the margin of receiving the
treatment at a given IV value $z$ under the monotonicity assumption. Similar to
$\tau^{\textup{\PRTE}} $, it is also a ``local" causal effect on the 
subpopulations
partitioned by the joint potential treatment status under different IV values.
In our application, $\tau^{\textup{\MTE}}(z)$ is the causal effect on the
children from the families who would have only one child at a given policy
intensity $z$ but would have more children with less intensity. In other words,
for each given intensity level $z$, $\tau^{\textup{\MTE}}(z)$ is the causal
effect on the children from the families who would change their fertility
decision to obey the OCP just at that level. The $\tau^{\textup{\MTE}}(z)$ over 
the range of the IV
give a complete picture of the heterogeneous treatment effects in the study
population.

\subsection{Local instrumental variable and principal stratification}

\citet{Heckman1999} derived nonparametric identification formulas for the three
causal estimands, which involve estimating the partial derivative of the
conditional expectation of the outcome given covariates and the estimated
propensity scores. \citet{Carneiro2011, Carneiro2017} used the sample analogues
of the partial derivative in the estimation. However, it is difficult to
quantify the uncertainty of these estimators as well as to accommodate
clustered data. Below we adopt a selection model representation of the problem, 
which naturally allows for flexible Bayesian inference based on hierarchical 
models. The key
to this representation is an intrinsic connection between the local IV approach
\citep{Heckman1999} and principal stratification 
\citep{Frangakis2002principal}, as illustrated below.

Principal stratification is a general framework for adjusting post-treatment
intermediate variables in causal inference, which extends the IV approach to
noncompliance by \cite{Angrist1996}. In the context of IV, the intermediate
variable is the treatment status, and a principal stratum is defined as the
joint potential values of the treatment under all possible values of the IV:
$\mathcal{T}_i  =  \{T_{i}(z): z\in[z_{\min},z_{\max}]\}$. The key insight
is that the principal stratum, by definition, is not affected by the observed
value of the IV and thus can be viewed as a latent pretreatment variable
\citep{Frangakis2002principal}. Therefore, one can define the principal causal
effects as comparisons of potential outcomes conditioning on one principal
stratum or combination of several principal strata. Both PRTE and MTE are
special cases of principal causal effects. Most applications of principal
stratification focused on a binary IV with a few exceptions
\citep[e.g.][]{jin2008principal,bartolucci2011copulas,schwartz2011bayesian}.
The main challenge to a continuous IV is that there can be infinitely many
possible principal stratum $\mathcal{T}_i$, rendering modelling and estimation
difficult.

Fortunately, we can reduce the principal stratum to a scalar. Under Assumptions
\ref{Unconfoundedness}--\ref{monotonicity}, \citet{Vytlacil2002} showed that
the potential treatment status $T_i(z)$ follows a latent selection model.
Namely, Assumptions \ref{Unconfoundedness}--\ref{monotonicity} are equivalent
to the following assumption.

\begin{assumption}\label{as:latent_sel} There exists a random variable $S_{i}$
	and a monotone non-degenerate function $v(\cdot)$ such that the intermediate
	treatment status $\{  T_{i}(z): z\in[z_{\min},z_{\max}] \}$ can be written 
	as
	$T_{i}(z)=\textup{\bf 1}_{v(z)\geq S_{i}}$, with $Z_{i}\ind
	S_{i}| X_{i} .$
\end{assumption}

In Assumption \ref{as:latent_sel}, $S_{i}$ is an unobserved threshold that
determines the treatment status of individual $i$. In our application, it
describes the latent ``utility'' of violating the OCP for individual $i$. A 
larger value of $S_{i}$ means that individual $i$
needs a larger encouragement $v(z)$ to obey the OCP. The selection model
representation in Assumption \ref{as:latent_sel} allows for flexible modeling
and inference strategies. See \citet{kline2019heckits} for more discussions on
the numerical equivalence between Assumption
\ref{Unconfoundedness}--\ref{monotonicity} and Assumption \ref{as:latent_sel}.

In Assumption \ref{as:latent_sel}, $v(z)$ is monotone in $z$. Without loss of 
generality, we set $v(z)=z$ because we can apply a monotone transformation of 
$v(z)$ and $S_{i}$ simultaneously. See \citet{Vytlacil2002} for more technical 
discussions. Therefore, $T_{i}(z)=\textbf{1}_{z\geq S_{i}}$ is a step function 
with respect of $z$ and its shape is determined by the random variable $S_i$. 
We can then use $S_i$ to characterize the entire vector of the potential values 
of the treatment $\mathcal{T}_i$, and thus will call $S_i$ the principal 
stratum hereafter. Based on the selection model representation in Assumption 
\ref{as:latent_sel}, the MTE estimand in \eqref{MTE_average} is equivalent to 
the following principal causal effect:
\begin{eqnarray}
\label{MTEre}
\tau^{\MTE}(s)& = & E\{Y_{i}(1)-Y_{i}(0)\mid S_{i}=s\}.
\end{eqnarray}

We can express the ATT and PRTE estimands as weighted averages of
$\tau^{\MTE}(s)$ over a range of principal strata. Specifically, averaging
$\tau^{\MTE}(s)$ over the distribution of $S_i$ for treated units leads to
$\tau^{\textup{\TT}}$:
\begin{gather}
\label{MTEtoATT}
\tau^{\textup{\TT}}
=\int_{-\infty}^{\infty}\tau^{\MTE}(s)F_{S}(\textup{d}s\mid T=1);
\end{gather}
averaging
$\tau^{\MTE}(s)$ over the distribution of $S_i$ between $[z_{\min},z_{\max}]$ 
leads to $\tau^{\textup{\PRTE}}$:
\begin{gather}
\label{MTEtoPRTE}
\tau^{\textup{\PRTE}}=\int_{z_{\min}}^{z_{\max}} \tau^{\MTE}(s)
F_{S}(\textup{d}s\mid z_{\min}\leq S \leq z_{\max}).
\end{gather}
In our context, the treated units are those with principal stratum $S_{i}$
smaller than the observed policy intensity. Importantly, \eqref{MTEtoATT} 
involves
the $\tau^{\MTE}(s)$ values for the principal strata below the minimal
intensity $z_{\min}$. However, because families with $S_i<z_{\min}$ or
$S_i>z_{\max}$ would not change their fertility decisions regardless of the
policy intensity (these are called the always-takers and never-takers by
\cite{Angrist1996}), the principal stratum $S_i$ of these families is not
unique even if we know the whole joint potential values $\mathcal{T}_i$. We
mitigate this complication by imposing a model for principal strata $S_i$
conditioning on covariates, based on which we can impute the individual
principal stratum membership $S_{i}$ for the always-takers or never-takers. A 
similar strategy was adopted in \citet{glickman2000derivation}.

\section{Bayesian hierarchical selection  and outcome models} 
\label{sec:Bayesian}	
\subsection{General structure of Bayesian inference}

The above connection between the local IV approach and principal stratification 
allows us to employ a flexible Bayesian modelling strategy to infer the causal 
estimands. Specifically, for each unit $i$, the complete data are 
$\{Y_{i}(1),Y_{i}(0), S_i, X_{i},Z_{i}\}$, where $S_i$ is equivalent to 
$\mathcal{T}_{i}$ under Assumptions \ref{Unconfoundedness}--\ref{ER}. The 
causal estimands are functions of the complete data, and thus inferring the 
causal effects depends on the complete data likelihood. Let 
$\Pr\{Y_{i}(1),Y_{i}(0), S_{i}, X_{i},Z_{i}\mid \theta\}$ denote the joint 
probability density function of the random variables for unit $i$ governed by 
parameters $\theta = (\zeta, \phi, \psi)$. We factorize the complete-data 
likelihood into three parts:
\begin{multline}
\label{stepa}
\Pr\{S_{i},Y_{i}(1),Y_{i}(0),X_{i},Z_{i} \mid \theta\}\\
=\Pr\{Y_{i}(1),Y_{i}(0) \mid S_{i},X_{i},\zeta \}
\times \Pr(S_{i} \mid X_{i},\phi)\times \Pr(X_{i},Z_{i} \mid \psi).
\end{multline}
In \eqref{stepa}, $\zeta$ denotes the parameters for the model of the potential 
outcomes, $\phi$ denotes the parameters for the model of the principal strata, 
and $\psi$ denotes the parameters for the distribution of the covariates and 
the IV. Following the common practice in the literature, we assume the three 
sets of parameters $(\zeta,\phi,\psi)$ are distinct and \emph{a priori} 
independent.

Based on the factorization in \eqref{stepa}, we can further refine the
definition of MTE, allowing it to be conditional on $X_{i}$:
\begin{eqnarray}
\label{MTEcondi}
\tau^{\MTE}(s,x)&=&E\{Y_{i}(1)- Y_{i}(0) \mid S_{i}=s,X_{i}=x \}.
\end{eqnarray}
Averaging $\tau^{\MTE}(s,x)$ over the distribution of $X_i $
conditional on $S_i = s$ gives the $\tau^{\MTE}(s)$ in \eqref{MTEre}.

The potential outcome $Y^{\textup{mis}}_i=Y_{i}(1-T_{i})$ and the principal
stratum $S_{i}$ are not observed for any unit. From a Bayesian perspective,
these unobserved values are no different from unknown model parameters, both of
which are unobserved random variables that we need to draw posterior inference
on \citep{Rubin1978}. Specifically, we will simulate the posterior distribution
of $\theta$, and impute the missing values $(Y^{\textup{mis}}_i, S_i)_{i=1}^N$
from their posterior predictive distributions conditional on the observed data
and $\theta$. We can then perform posterior inference of the causal effects
based on the posterior samples of $\theta$ and $( Y^{\textup{mis}}_i, S_i
)_{i=1}^N $.

\subsection{Models and posterior inference}

The factorization in \eqref{stepa} suggests that to infer causal effects we
need to specify three models: (a) a model for principal strata conditional on
the covariates $\Pr(S_{i} \mid X_{i} ,\phi)$, (b) a model for the
potential outcomes conditional on the covariates and principal stratum
$\Pr(Y_{i} \mid S_{i},X_{i},\phi)$, and (c) the joint distribution of the
observed covariates and IV $\Pr(X_{i},Z_{i} \mid \psi)$. In Bayesian
inference, we usually condition on the empirical distribution of the covariates
instead of modelling the joint distribution \citep{Ding2018review}, and thus
below we will focus on the first two models.

First, for the continuous principal stratum $S_i$, we postulate a hierarchical
model that accounts for clustering:
\begin{eqnarray}\label{eq:ps_model}
S_{i}&\sim& \mathcal{N}(\beta_{S}'X_{i}+r_{C_{i}},\sigma^{2}), \nonumber\\
T_{i}(z)&=&\textbf{1}_{z\geq S_{i}},
\end{eqnarray}
where $r_{C_{i}} \overset{\text{i.i.d.}}{\sim} \mathcal{N}(0,\tau_{S}^{2})$ for
$C_{i}=1,2,\cdots,G$, are the random effects capturing the correlation
structure of the principal stratum membership $S_{i}$ of the units within the
same province $C_{i}$. In our application, $r_{C_{i}}$ can be interpreted as the
latent resistance to the OCP in province $C_i$. This model implies the
correlation structure for the treatment assignment mechanism $\Pr(T_{i}=1 \mid
Z_{i},X_{i})$, which is equivalent to a Probit model with Normally
distributed random effects:
\begin{equation}\label{eq:probit}
\Pr(T_{i}=1\mid
Z_{i},X_{i})=\Phi\{(Z_{i}-\beta_{S}'X_{i}-r_{C_{i}})/\sigma\},
\end{equation}
where $\Phi(\cdot)$ is the cumulative distribution function for the standard
Normal distribution. Here we restrict the coefficient of $Z_{i}$ to be $1$ but
allow for an unknown variance of $\varepsilon_i$. This parametrization differs
from the standard Probit model but follows more closely the latent index
representation in Assumption \ref{as:latent_sel}.

Second, for the ordinal potential outcomes, we postulate a proportional odds
model with a cumulative logit link \citep{Agresti2003}: for $t=0,1$ and
$k=1,2,\cdots K-1$,
\begin{eqnarray}
\textup{logit}\{\Pr(Y_{i}(t)\leq k\mid X_{i},S_{i})\} =
\alpha_{t,k}+\beta_{t}' X_{i}+\gamma_{t}S_{i}+\nu_{t, C_{i}}, \quad \text{
	with }
\alpha_{t,k} <  \alpha_{t,k+1},  \label{eq:outcome_model}
\end{eqnarray}
where $\nu_{t,C_{i}} \overset{\text{i.i.d.}}{\sim} \mathcal{N}(0,\tau_{t}^{2})
$ for  $C_{i}=1,2,\cdots,G$, capture the correlation structure of the potential
outcomes within a province. In our application, $\nu_{t,C_{i}}$ can be
interpreted as the latent psychological characteristics in province $C_i$.
Model \eqref{eq:outcome_model} assumes that each potential outcome has its own
increasing intercepts $\alpha_{t,k}$'s for $t=0,1$. The outcome model
\eqref{eq:outcome_model} differs from the classical selection model
\citep{heckman1979sample}: The former models the conditional distribution of
$Y_i(t)$ given $S_i$, whereas the latter models the joint distribution of the
treatment assignment and the outcome model.
The parametrization in \eqref{eq:outcome_model} offers more convenient 
inference for the MTE.
We do not impose a joint model for
$Y_i(1)$ and $Y_i(0)$ because the data contain no information about their
association. This does not pose a problem for inferring the $\tau$ estimands
because they depend only on the marginal distributions of the potential
outcomes.

We impose standard weakly-informative priors for the parameters. For the
regression coefficients $\beta$, $\alpha$ and $\gamma$, we impose the diffuse
Normal priors $\beta_{d} \sim \mathcal{N}(0,100\times I_p)$, $\beta_{t}\sim
\mathcal{N}(0,100\times I_p)$ and $\alpha_{tk}\sim \mathcal{N}(0,100\times
I_{K-1})$, and $\gamma_{t}\sim \mathcal{N}(0,100)$; for the variance of the
Probit model, we impose a flat prior $\pi(\sigma^{2}) \propto 1/\sigma^{2}$;
for the standard deviations of the random effects $\tau_{S}$ and $\tau_{t}$, we
impose the half-Cauchy priors $	\pi(\tau_{S} ) \propto \{  1+(\tau_{S}/A)^{2}
\}^{-1}\textbf{1}_{\tau_{S}\geq 0}$ and  $\pi(\tau_{t}) \propto \{
1+(\tau_{t}/A)^{2} \}^{-1}\textbf{1}_{\tau_{t}\geq 0}$ with the scale parameter
$A=25$ \citep{gelman2006prior}. Given the models and the prior distributions of
the model parameters, we can obtain the posterior distribution of
$\zeta=\{\alpha_{t,k},\beta_{t},\gamma_{t},\tau_{t}^{2} : t=0,1,k=1,\ldots K-1
\} $. We then derive the posterior distributions of
$\tau^{\textup{\MTE}}(s,x)$. These are all functions of $\zeta$ based on the
hierarchical outcome model. Finally, we average these conditional effects over
the empirical distribution of $\{S_{i},X_{i},C_{i}\}$ to obtain the
unconditional effects. For example, given a posterior draw of the parameters
$\zeta$, the province random effect $v_{t,C_{i}}$, and principal stratum 
$S_{i}$, we
can obtain a posterior draw of
$\tau^{\textup{\TT}}$ and $\tau^{\textup{\PRTE}}$ from
\begin{eqnarray}
\label{ATTexp}
\tau^{\textup{\TT}}&=& \sum_{i=1}^{N}T_{i}\tau^{\MTE}(S_{i},X_{i};
\zeta,v_{t,C_{i}})
\Big / \sum_{i=1}^{N}T_{i}.\\
\label{LATEexp}
\tau^{\textup{\PRTE}}&=&
\sum_{i=1}^{N}\delta_{i}\tau^{\MTE}(S_{i},X_{i};\zeta,v_{t,C_{i}}) \Big /
\sum_{i=1}^{N}\delta_{i},
\mbox{ with }  \delta_{i}=\textbf{1}_{ z_{\min}\leq S_{i}\leq z_{\max} }.
\end{eqnarray}
This yields the posterior distributions of $\tau^{\textup{\TT}}$ and $
\tau^{\textup{\PRTE}}$. In  \eqref{ATTexp} and \eqref{LATEexp}, we emphasize
that the conditional effects depend on the parameter $\zeta$ and province
random effects $v_{t,C_{i}}$.

\section{Empirical application} \label{sec:app}

\subsection{The data and preliminary analysis}
We now provide more information about the data. The CFPS is a comprensive
longitudinal household survey representing 95\% population in China. We only
analyze the data from the first national wave in 2010, so that each household
only
appears once in the questionaire. The survey
sample covers approximately 14,000
households from 25 provinces. Following \cite{Wu2014},  we exclude the samples 
from provinces such as
Ninxia, Gansu, Xinjiang, and Yunnan, which include a large proportion
of minority ethnicity and hence avoid the OCP restriction.

From the 2010 wave of CFPS we
obtain the subsample of children born after 1979, the year when the OCP was
first imposed, with age ranging  from 16 to 31. Pretreatment
covariates  include age, ethnicity (1 for the Han ethnicity, 0 otherwise),
paternal and maternal education attainment measured in years, family
income,  parents marriage status (whether divorced or not), and parental
age at the child birth. We stratify on
the area (urban or rural) and gender (female or male) and split
the sample into four subgroups: rural females, rural males, urban females, and
urban males, with a sample size of 1747, 1708, 407, and 416, respectively.
Within these subgroups, the numbers of only children are 147, 283, 254, and
296, respectively. For families violating the OCP, we only keep the oldest
child within a household. Clearly, the urban area has a much higher proportion
of only
children than the rural areas: the proportions of the households with
only children are 12.46\% and 66.89\% in rural and urban areas, respectively. We
use three psychological measures as
outcome variables: self-confidence, degree of anxiety and desperation. During
the survey, investigators asked the interviewees for the frequency of
experiencing these emotions, and then transformed the frequency into the Likert
scale \citep{Likert1932}. The outcomes take discrete values from 1 to 5, with a
larger value indicating a better psychological condition.

In Table \ref{tab:covariates}, the summary statistics of the observed
covariates between the families with one child (i.e. treated) and the families
with more than one child (i.e. control) reveal notable differences in the
pretreatment variables between the treatment and control groups. In general,
parents of only children had more educational attainment, but had a higher 
divorce rate. Also, only
children were younger on average, which is partly due to our choice of keeping 
only the oldest child
from a family with multiple children. This choice is to eliminate the 
difference in the birth
order in the study sample, which may have a lasting effect on personality 
\citep{rohrer2015examining}. Moreover, families with only children on average 
had higher household income. The Han ethnicity had a higher proportion of only 
children than minor ethnicities; this is expected because
the OCP was not enforced among the ethnical minorities. In
addition, we control for the number of siblings in the households with
more than one child. For females, we also check whether she has a younger
brother, which could potentially influence her psychological
health \citep{chu2007effects}. Specifically, we create a dummy variable
that equals zero for the girls who were only children or only had
female siblings and equals one for the girls who had a younger brother.
However, simple $t$-tests do not reveal any significant differences in
the outcomes between the treatment and control groups.

\begin{table}
	\caption{\label{tab:covariates}Baseline characteristics of the whole sample,
		households with only children, and households with multiple children. 
		Standard deviations are presented in the parentheses. }
	\centering
	\begin{threeparttable}	
		\begin{tabular}{lccc}
			\hline
			\hline
			& All samples &Only child  & With siblings
			\\			
			&$N=4278$&$N_{1}=980$&$N_{0}=3298$\\
			\hline
			
			\textit{Family background}&&&\\
			Maternal education (years) &$5.05\ (4.57)$ &$7.93\ (4.28)$&$4.19 \ 
			(4.29)$\\
			Paternal education (years) &$6.92\ (4.39)$&$8.71\  (3.99)$&$6.39 \
			(4.37)$\\
			Maternal age at birth (years) &$25.5\ (4.34)$ &$25.1\  
			(3.44)$&$25.6 \
			(4.57)$\\
			Paternal age at birth (years) &$27.4\  (4.92)$&$26.8\  
			(3.82)$&$27.6 \
			(5.20)$\\
			Family annual income (CNY\footnote{})
			&$45100\  (165790)$ &$59025\  (87050)$ &$40965 \
			(182570)$\\
			Divorce &$1.49\% \  (12.1\%)$&$2.95\%\  (16.9\%)$&$1.06\% \
			(10.2\%)$\\
			\\
			\textit{Individual information}&&&\\
			Age (years)  &$26.5\ (3.49)$&$25.0\  (3.38)$&$27.2\ (3.52)$\\
			Number of siblings  &$1.35\  (1.17)$&$0\  (0)$&$1.74\ (1.04)$\\
			Proportion of majority ethnicity &$90.6\% \
			(29.1\%)$&$95.6\%\ (20.4\%)$&$89.1\%\ (31.1\%)$\\
			\\
			\textit{Outcome information}&&&\\
			Confidence measure &$4.00 \ (0.95)$&$ 3.97\  (0.91)$& $4.01\  
			(0.97)$\\
			Anxiety measure & $4.16\  (1.06)$& $4.14 \ (1.06)$& $4.17\  
			(1.07)$\\
			Desperate measure &$4.26\  (1.03)$& $4.22\ (1.03)$& $4.27 \ 
			(1.03)$\\
			\hline
		\end{tabular}
		\begin{tablenotes}
			\item[\textdagger] Annual income in the Chinese currency (CNY) is 
			for the year when the household was surveyed.
		\end{tablenotes}
	\end{threeparttable}
\end{table}

\begin{figure}[h]
	\centering
	\includegraphics[width= 0.8\textwidth]{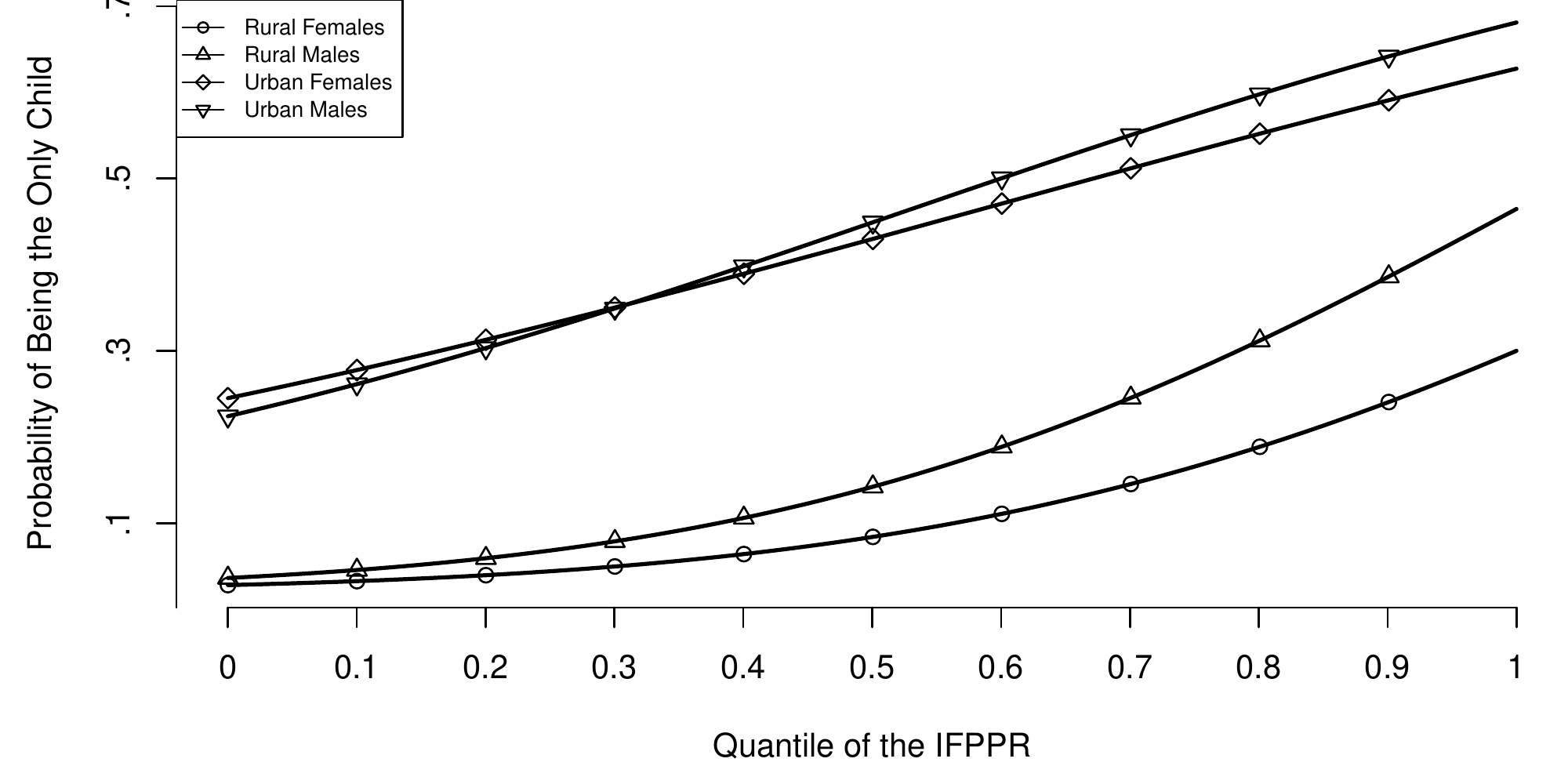}
	\caption{Probability of being treated (i.e. being an only child) as a 
	function of the IFPPR quantiles}
	\label{Firststage}
\end{figure}

Our IV is the implementation intensity of the OCP, measured by the Indicator of
Family Planning Policy Resistance (IFPPR) \citep{Attane2002}, ranging from 0 to
140. This index is an aggregated measure for the implementation intensity
of OCP in the period of 1980s and hence does not vary with time. As discussed in
Section \ref{sec:methods}, Assumption 1 (randomization of the IV) and
Assumption 3 (monotonicity) is reasonable in our application.
To examine Assumption \ref{as:nontrivial},  Figure \ref{Firststage} shows  the 
fitted probabilities of being an only child in a family as a function
of the quantile of the IV, which are
predicted from the hierarchical model \eqref{eq:ps_model} with covariates fixed
at their means and the random effects fixed at zero.

Figure \ref{Firststage} shows a clear increasing trend of being an only
child as the IFPPR goes up in all four subpopulations, bolstering the
plausibility of Assumption \ref{as:nontrivial}. We also see the heterogeneity
in the treatment assignment mechanism across the four subpopulations. For both 
females
and males, the individuals from the urban areas have a larger probability of
being  only children compared to those from the rural areas. This pattern can
be attributed to the fact that urban residents were more inclined to obey
government regulations because (a) the consequences of violating the OCP were
usually more severe in the urban areas, and (b) urban residents were less
likely to hold the traditional thinking of ``more children are better'' that
was quite common in the rural areas. There is also a marked difference between
females and males, especially in the rural areas. Girls had a lower probability
of being  only children. This pattern could be attributed to the
traditional perception in China that ``boys are superior to girls'', namely, 
the families
who had a girl as their first child would attempt to have a second child in
hope of having a boy, even at the cost of violating the OCP. We observe a
larger gender gap in the rural areas where such a perception was particularly
prevalent. Another explanation for the discrepancy between urban
and rural areas is the policy change. In mid 1980s, the Chinese government 
gradually relaxed the
restriction in rural ares, allowing families to have a second child if the
first one was a daughter \citep{banister1991china,huang2016fertility}. As a 
result, the
proportion of female only children was lower in rural areas.

\subsection{Marginal and average treatment effects}
We applied our local IV approach with the Bayesian models \eqref{eq:ps_model} 
and \eqref{eq:outcome_model} and priors in Section \ref{sec:Bayesian} to the 
CFPS data. We conducted separate analyses within each of the four 
subpopulations. We used \texttt{JAGS} \citep{plummer2003jags} to simulate the 
posterior distributions. For each analysis, we simulated three Markov chains 
with different starting values and 50,000 iterations for each chain, discarding 
the first half as burn-in and thinning the chains for every 25 iterations. The 
remaining 3,000 draws were used to approximate the posterior distributions. The 
Gelman--Rubin statistic \citep{Gelman1992} for all parameters are below 1.1 
indicating the good mixing of the Markov chains.

\begin{figure}[H]
	\centering
	\includegraphics[width=1\textwidth]{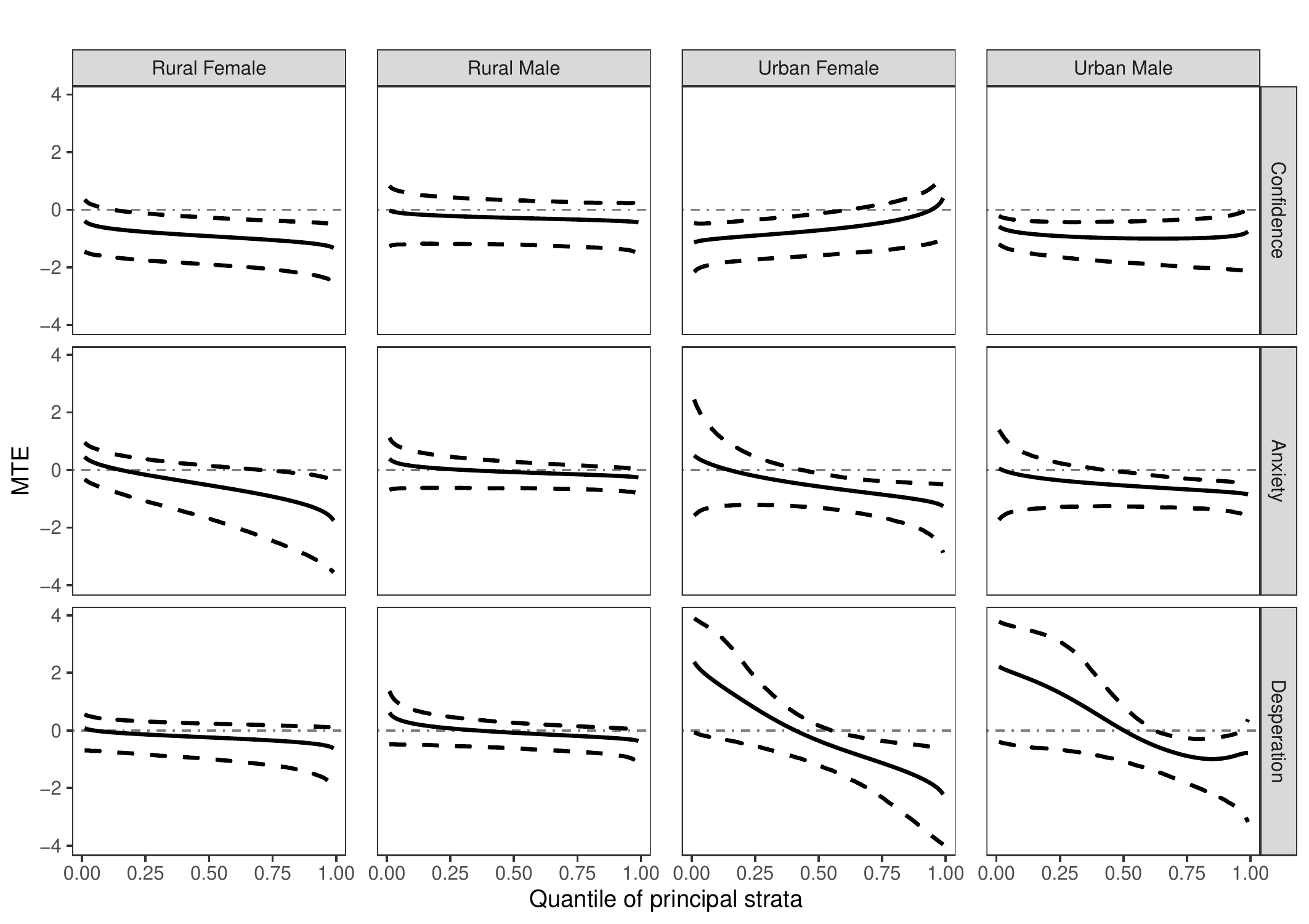}
	\caption{\label{MTEResult}Posterior means and 95\% credible intervals of
		$\tau^{\textup{\MTE}}(s, x)$ against  principal strata $s$.
		X-axis represents the quantile of the principal strata $S_{i}$ after
		normalization. A larger value in the x-axis means that the individual 
		is from a family that is more resistant to the OCP.}
\end{figure}

Figure \ref{MTEResult} shows the posterior means and credible intervals of the 
marginal treatment effects $\tau^{\textup{\MTE}}(s,x)$  against  the principal 
stratum $S_i$, for the three outcomes. To compare across different subgroups 
and outcomes, we transform the principal stratum value into the quantile of its 
posterior distribution based on the imputed $S_{i}$ in the sample. We use a 
formula from \citet[][page 133]{ju2010criteria} to obtain the analytic 
expression of $\tau^{\textup{\MTE}}(s,x)$  from the hierarchical outcome model 
\eqref{eq:outcome_model} as follows:
\begin{eqnarray*}
	\tau_i(s,x)&=&E\{Y_{i}(1)-Y_{i}(0)\mid X_{i}=x,S_{i}=s\}\\
	&=&\sum_{k=1}^{K-1}\{\textup{sig}(\alpha_{0,k}+\beta_{0}'x+\gamma_{0}s+v_{0,C_i})-\textup{sig}(\alpha_{1,k}+\beta_{1}'x+\gamma_{1}s+v_{1,C_i})\},
\end{eqnarray*}
where $\textup{sig}(x)=1/\{1+\textup{exp}(-x)\}$ is the sigmoid function. We 
then evaluate the posterior means and credible intervals at a grid of values of 
the quantiles of the principal stratum $S_i$, again setting the covariates at 
their means and the random effects at zero.

For the confidence measure, a range of $\tau^{\MTE}(s,x)$ lie below zero for 
the subgroups of rural females, urban females, and urban males, indicating 
negative treatment effects across  principal strata. The males from  urban 
areas are more likely to have a worse confidence measure, compared with males 
from  rural areas. The patterns of $\tau^{\MTE}(x,s)$ are similar for the 
anxiety and desperation measures. We observe a decreasing trend in the 
treatment effect as the principal stratification $S_i$ increases. This trend is 
more apparent for the urban subgroups. Recall that $\tau^{\textup{\MTE}}(x,s)$ 
represents how the treatment effect of being an only child varies across 
principal strata, which quantify families's reluctancy of bearing only one 
child. That is, for those from families that were more resistant to the OCP, 
the effect of being an only child is more negative. For instance, in the bottom 
right corner of Figure \ref{MTEResult}, the effect on desperation measure for 
urban males flips its sign from approximately $2$ to $-1$ as the reluctance to 
obey the policy increases, taking up about $40\%$ and $-20\%$ compared to the 
range of the outcome. A possible explanation is that the families with stronger 
preference for more children did not prepare well to bear only one child and 
thus provided a less nurturing environment for their children. In contrast, the 
decreasing trend is less noticeable for the rural subgroups, such as the 
desperation measure for rural males in the bottom left of Figure 
\ref{MTEResult}. Indeed, in the rural subgroups the effects of being an only 
child vary little across the families' degree of resistance to the OCP.

\begin{figure}
	\centering
	\includegraphics[scale=0.65]{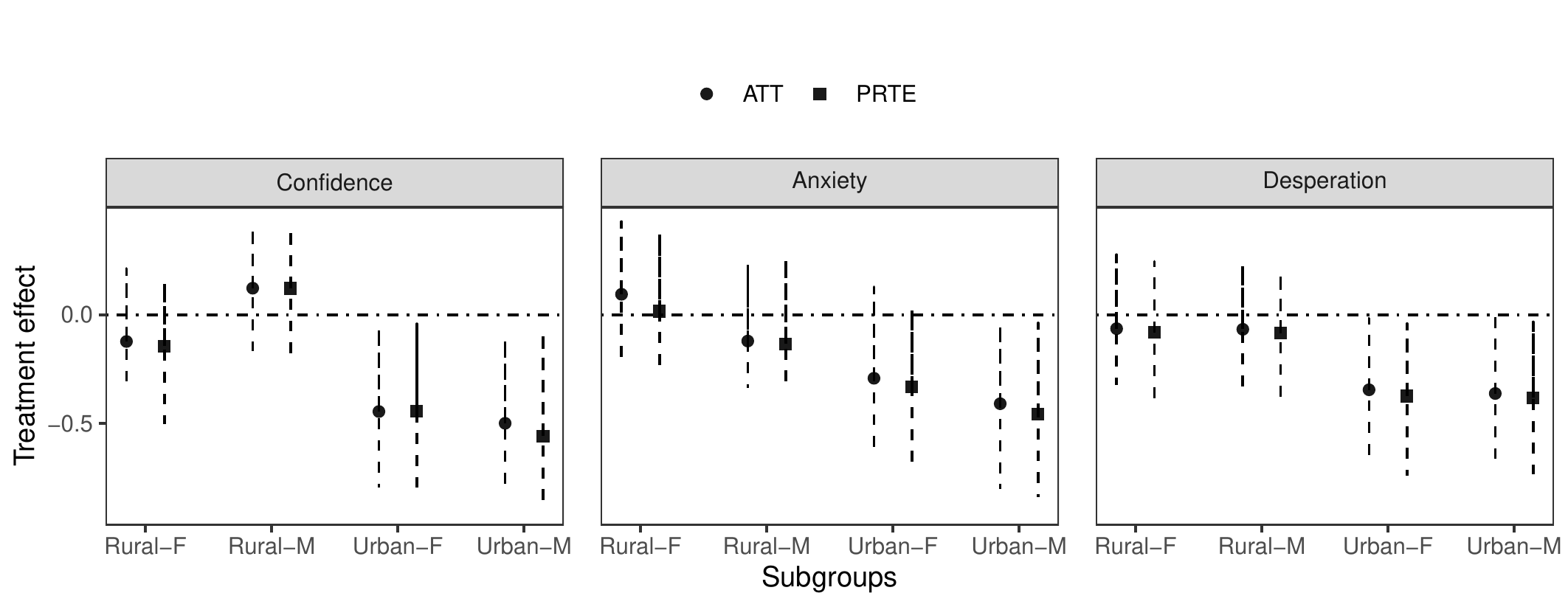}
	\caption{\label{overall_plot}Posterior means and 95\% credible
		Intervals of $\tau^{\PRTE}$ and $\tau^{\TT}$ . The $\circ$'s are the  
		posterior
		means   of the treated effects on the treated and the $\square$'s are 
		the posterior means of   the policy related treatment
		effects. The vertical dashed lines are the credible intervals.}
\end{figure}

Based on formulas (\ref{ATTexp}) and (\ref{LATEexp}), we can infer
$\tau^{\textup{\TT}}$ and $\tau^{\textup{\PRTE}}$ based on their posterior 
means and 95\%
credible intervals, which are displayed in Figure \ref{overall_plot}.
Overall, $\tau^{\textup{\TT}}$ and $\tau^{\textup{\PRTE}}$ are very similar for
each subgroup and outcome, with $\tau^{\textup{\TT}}$ being slightly smaller.
For all three outcome measures, in urban areas only children have on
average $0.3-0.5$ (under the outcome scale of 1--5) smaller values than the 
children with
siblings, translating into a statistically significant $7.5\%-12.5\%$ decrease
in self-reported psychological health. The effects on the rural individuals are
much smaller and inconclusive. This pattern is consistent with the results from
the marginal effects $\tau^{\textup{\MTE}}$.  One  explanation is
that the children grew up in rural areas had less restriction to
communicate with their peers in the same community, such as a village. The 
companion of other children in the same cohort might substitute the effect of 
siblings, while the children in urban area might not have such
opportunities.

It is worth noting that $\tau^{\textup{\TT}}$ averages over all families with 
only children regardless of the OCP, whereas $\tau^{\textup{\PRTE}}$ averages 
over families who had only children as the consequence of the OCP. The 
similarity between $\tau^{\textup{\TT}}$ and $\tau^{\textup{\PRTE}}$ is due to 
the large overlap of the target populations of $\tau^{\textup{\TT}}$  and 
$\tau^{\textup{\PRTE}}$. Namely, most families having only one child, 
corresponding to the target population of $\tau^{\textup{\TT}}$, would not have 
only one child at the lowest policy intensity, which belong to the target 
population of $\tau^{\textup{\PRTE}}$. Indeed we found that the imputed 
principal strata of the majority (between 55\% to 75\%) of the families lie in 
the range of the observed values of the IV in all four subgroups. Therefore, 
the policy affected most families, in the sense that most families would have 
more than one child at the lowest policy intensity but have only one child at 
the highest policy intensity.

In summary, our analysis suggests that being an only child has a negative
impact on psychological health, especially for those from the urban areas.
Specifically, being an only child decreases approximately 12\%, 7.5\% and
10\%  in the confidence, anxiety and desperation measures, respectively. One
possible explanation of the negative effect is that only children might be 
subject to more pressure from parents, which
influences psychological health, giving rise to an individual being less
confident and more likely to feel anxious and desperate. The individuals from 
urban
areas, especially urban males, might receive higher expectations from  their
families compared with those from rural areas. This leads to a worse impact of 
being
an only child on the psychological health of the urban subgroup.

\subsection{Comparison with alternative methods} \label{sec:compare}

We now compare the proposed local IV method with two
alternative standard methods. The first method is direct regression adjustment 
based on ordinary least squares
(OLS), where we fit a regression model of the outcome on the treatment and
centered pretreatment variables with cluster-robust standard errors, $E(Y_i\mid
T_i, X_i)=\beta_0+\beta_t T_i+ \beta_x X_i+\beta_{tx}T_iX_i$, and take the
estimated coefficient $\beta_t$ to estimate the ATT. However, OLS cannot
estimate the MTE or PRTE which are local effects defined by the IV. Moreover, 
OLS
relies on the unconfoundedness assumption, that is, there is no unmeasured
confounding beyond the observed pretreatment variables, which is unlikely to
hold given the limited covariate information in the CPFS data.

The second method is the standard two-stage least squares (2SLS) for IV 
analysis. In the first stage, we use a regression model of $T_{i}$ on $X_{i}$ 
and $Z_{i}$, and in the second stage, we regress the outcome $Y_{i}$ on $X_{i}$ 
and the fitted values of $T_i$ from the first stage to obtain the coefficient 
of the fitted $T$   as an estimate of the ATT. We use the standard errors 
robust to the cluster structure in the two stages. The 2SLS method is more 
comparable to the local IV method. However, while the 2SLS estimate has a clear 
causal interpretation similar to the ATT when the treatment effects are 
homogeneous, its interpretation becomes ambiguous when treatment effects are 
heterogeneous \citep{heckman2005structural}. In contrast, the local IV method 
can provide a comprehensive picture of the potentially heterogeneous treatment 
effects via the marginal treatment effects curves shown in Figure 
\ref{MTEResult}.

\begin{figure}[ht]
	\centering
	\includegraphics[width = \textwidth]{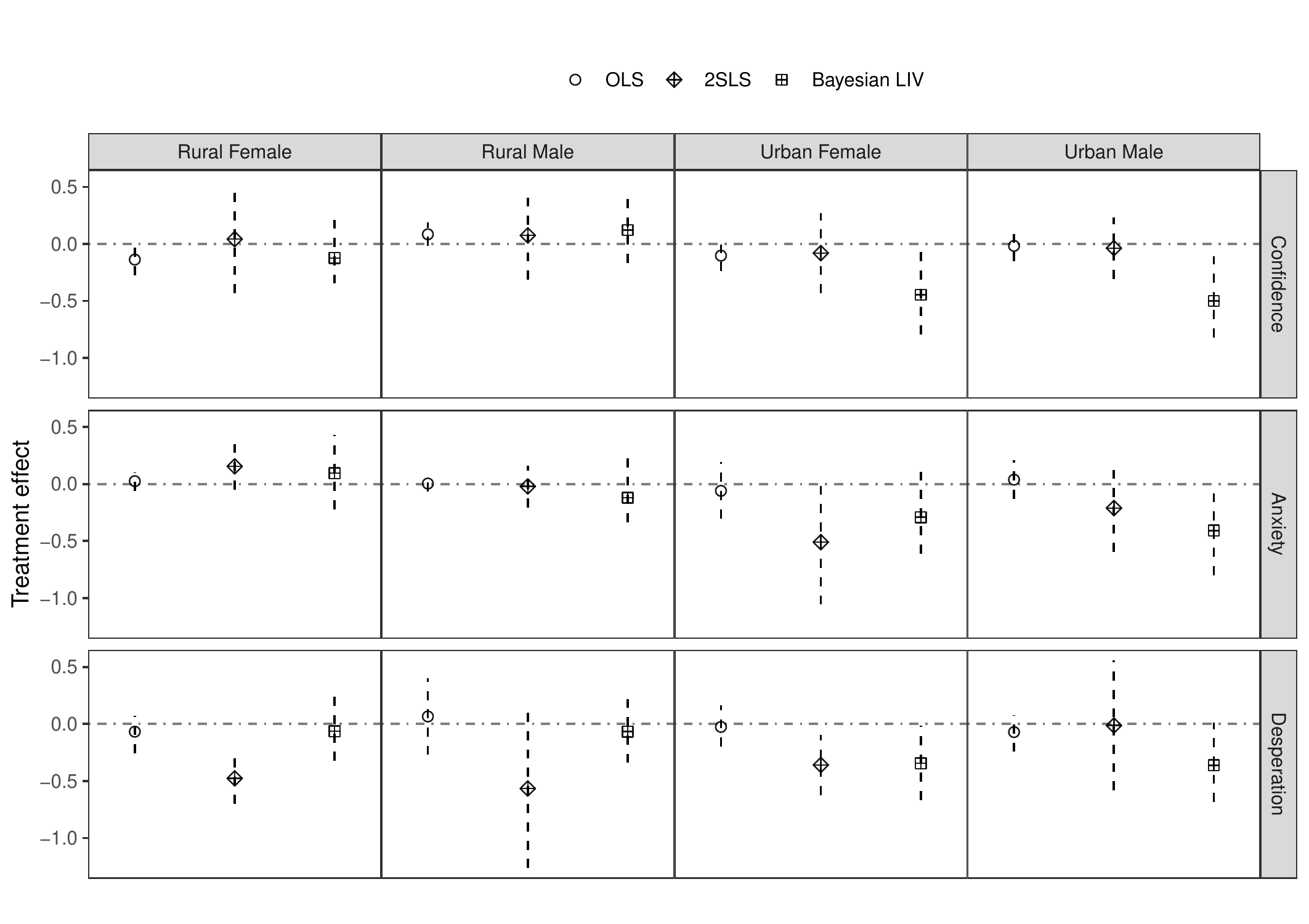}
	\caption{\label{OtherMethod}Comparison with   other
		methods: dashed lines represent confidence or credible intervals.}
\end{figure}

Figure \ref{OtherMethod} shows the point estimates and 95\% confidence
intervals of the ATT estimated by OLS, 2SLS and local IV. The OLS estimates
have much smaller standard errors than both IV methods, but the estimates are
concentrated around zero and fail to detect significant effect in any outcomes
and subgroups. Local IV and 2SLS  generally agree on the signs of the
effects, but local IV detects more significant effects with smaller standard
errors than 2SLS. More importantly, as shown in the MTE curves from local
IV, there is significant heterogeneity in treatment effects among urban
children, particularly in anxiety and depression; reporting only the average
effects as 2SLS and OLS would not capture the full picture of the heterogeneous
effects.

\subsection{Sensitivity analysis} \label{sec:SA}

Among the causal assumptions we made, exclusion restriction in Assumption 
\ref{ER} is the least justifiable by subjective knowledge. Therefore, in this 
subsection we conduct sensitivity analysis to examine the impact of the 
potential violation to exclusion restriction. To define the sensitivity 
parameter, we decompose the effect of the IV on the outcome into the component 
acting through the treatment and the remaining component:
\begin{eqnarray*}\label{directeffect}
	\tau^{\textup{direct}}&=&E\{Y_{i}(z_{\textup{max}}, 
	0)-Y_{i}(z_{\textup{min}}, 0)\},\\
	\tau^{\text{indirect}}&=&E\{Y_{i}(z_{\textup{max}}, 
	1)-Y_{i}(z_{\textup{max}}, 0)\},
\end{eqnarray*}
and define the sum of the two as:
$$
\tau^{\textup{total}} =
E\{Y_{i}(z_{\textup{max}}, 1)-Y_{i}( z_{\textup{min}}, 0)\}  =
\tau^{\textup{direct}}+	\tau^{\text{indirect}}.
$$
We call these three causal estimands as the direct, indirect and total effect 
for convenience, acknowledging that they are different from the ones in 
mediation analysis \citep{baron1986moderator,imai2010general}. If Assumption 
\ref{ER} holds, namely $Y(z, t)=Y(t)$, then the direct effect of the IV equals 
zero and thus $\tau^{\textup{total}}=\tau^{\textup{indirect}}$. Therefore, we 
use the ratio of $\tau^{\textup{direct}}$ and $\tau^{\textup{total}}$ as the 
sensitivity parameter $r=\tau^{\textup{direct}}/\tau^{\textup{total}}$. When 
Assumption \ref{ER} holds, $r=0$ and $|r|$ increases as the degree of violation 
to Assumption \ref{ER} increases.  If we assume the scale of the direct effect 
$|\tau^{\textup{direct}}|$ never exceeds that of the total effect 
$|\tau^{\textup{total}}|$, the maximum value of $|r|$ equals one.

To conduct sensitivity analysis with respect to exclusion restriction, we add 
an IV term into the outcome model \eqref{eq:outcome_model},
\begin{eqnarray*}\label{sensitivity_analysis}
	\textup{logit}\{\Pr(Y_{i}(z, t)\leq k\mid X_{i},S_{i},C_{i})\}
	= \alpha_{t,k}+\beta_{t}' X_{i}+\gamma_{t}S_{i}+\delta z+\nu_{t, C_{i}},
\end{eqnarray*}
where $\delta$ measures how the IV directly affects the outcome. The 
coefficient $\delta$ determines the sensitivity parameter $r$ based on   the 
formula
$$
r=\frac{\sum_{i=1}^{N}\sum_{k=1}^{K-1}\{\textup{sig}(\alpha_{0,k}+\beta_{0}'X_{i}+\gamma_{0}S_{i}+\delta
z_{\textup{min}})-\textup{sig}(\alpha_{0,k}+\beta_{0}'X_{i}+\gamma_{0}S_{i}+\delta
z_{\textup{max}})\} 
}{\sum_{i=1}^{N}\sum_{k=1}^{K-1}\{\textup{sig}(\alpha_{0,k}+\beta_{0}'X_{i}+\gamma_{0}S_{i}+\delta
 z_{\textup{min}})-\textup{sig}(\alpha_{1,k}+\beta_{1}'X_{i}+\gamma_{1}S_{i}+\delta
 z_{\textup{max}})\} }.$$
With a fixed value of $\delta$ and thus $r$, we can follow exactly the same 
procedure as before to estimate $\tau^{\TT}$ and $\tau^{\PRTE}$.

\begin{figure}[h]
	\centering
	\includegraphics[scale=0.68]{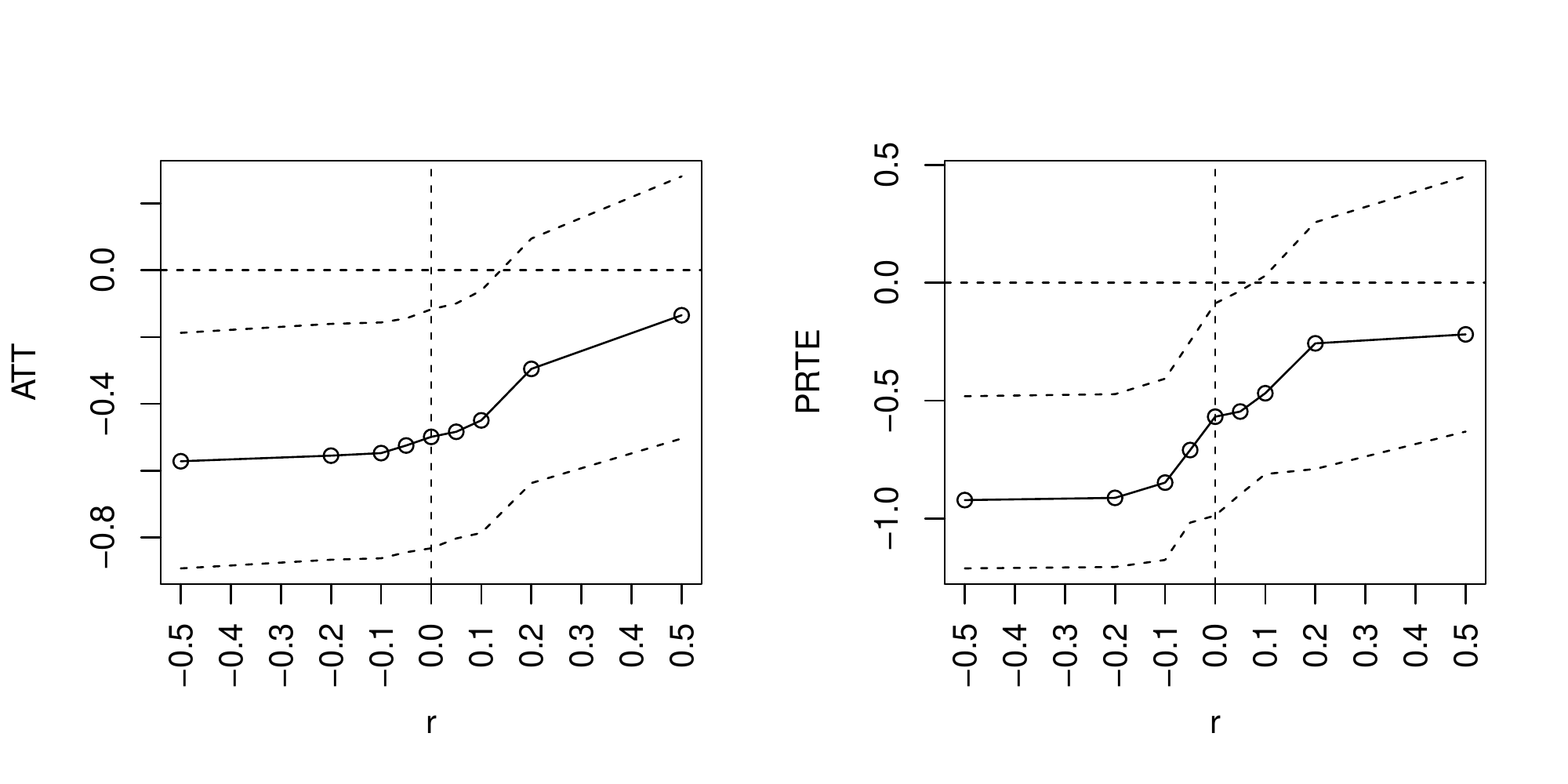}
	\caption{Posterior means and 95\% credible intervals of $\tau^{\PRTE}$ and
		$\tau^{\TT}$ against the sensitivity parameter $r$, for confidence 
		measure and
		urban males; $r=0$ when exclusion restriction holds.}.
	\label{fig:sa}
\end{figure}

We plot the estimated $\tau^{\TT}$ and $\tau^{\PRTE}$ against the sensitivity 
parameter $r$ in Figure \ref{fig:sa}, with $r$ on a grid ranging from $-0.5$ to 
$0.5$. As a result, we only examine the case where the direct effect from the 
IV takes up at most 50\% of total effect, which is a very large proportion in 
our application. For illustration purposes, we only display the effect on 
confidence measure of the urban males, which exhibits a strong negative effect 
under exclusion restriction. Figure \ref{fig:sa} shows an increasing trend 
between the treatment effect and sensitivity parameter $r$. This pattern is 
reasonable as the sensitivity parameter characterizes how much of the effect 
can be explained by the influence of  the IV  acting directly on the outcome 
without affecting the treatment. If $r$ is close to 1, a large proportion of 
the difference can be explained by the direct impact of the IV, which leads the 
indirect effect through the treatment changing from negative to zero. 
Therefore, our previous conclusions are more likely to incorrectly identify the 
negative effect when the sensitivity parameter $r$ is larger.

We examine the threshold below which the indirect effect remains significantly 
negative, and thus quantify the robustness of our previous estimates. The 
estimated effect remains significantly negative when $r$ is below 0.1. Namely, 
as long as the IV takes up less than 10\% change in the variation of the 
outcomes, the impact of being an only child on the confidence measure remains 
negative for urban males. In Figure \ref{fig:sa}, the sensitivity of 
$\tau^{\TT}$ and $\tau^{\PRTE}$ with respect to Assumption \ref{ER} share 
nearly identical trend against the sensitivity parameter. We relegate the 
sensitivity analysis results for the other measures and subgroups to the 
supplementary document.

\section{Simulations} \label{sec:simulations}
We further examine the operating characteristics of the proposed local IV
method and compare it with the OLS and 2SLS methods in simulated scenarios. We
independently simulate a covariate $X_{i}\sim \mathcal{N}(0,1)$ and an IV 
$Z_{i}\sim \mathcal{N}(0,1)$. We posit the following
latent threshold model for the treatment $T_{i}$:
\begin{eqnarray*}
	S_{i} &=& 0.5-0.5X_{i}+\varepsilon_{i},\\
	T_{i} &=& \textbf{1}_{Z_{i}\geq S_{i}},
\end{eqnarray*}
where $\varepsilon_{i}\sim \mathcal{N}(0,1)$. The latent threshold $S_{i}$ is a
function of the covariate, and an individual is treated when $Z_{i}$ exceeds
the threshold value. For the potential outcomes, we posit  linear models:
\begin{eqnarray*}
	Y_{i}(0)&=&b_{00}+b_{01}X_{i}+e_{i},\\
	Y_{i}(1)&=&b_{10}+b_{11}X_{i}+e_{i}+l_{i},
\end{eqnarray*}
We assume that $(e_{i},l_{i},\varepsilon_{i})$ follows multivariate Normal with
mean zero, variance one, and the following correlation structure
\begin{equation*}
\left(
\begin{array}{ccc}
1 & 0 & p\\
0& 1 & h\\
p & h & 1\\
\end{array} \right) .
\end{equation*}
The above model allows us to separate the
correlation between treatment $T_{i}$ and potential outcomes
$Y_{i}(1),Y_{i}(0)$ and the correlation between $T_{i}$ and treatment effect
$Y_{i}(1)-Y_{i}(0)$. Specifically, the correlation $p$ between errors $e_{i}$
and $\varepsilon_{i}$ controls the degree of confounding between the treatment
and the outcome, and the correlation $h$ between $l_{i}$ and $\varepsilon_{i}$
controls the degree of treatment effect heterogeneity with respect to the
threshold value $\varepsilon_{i}$. A larger $|h|$ corresponds to larger
heterogeneity, namely, the marginal treatment effect curve
$\tau^{\textup{MTE}}$ has larger variation across the principal stratum
$S_{i}$.

To be consistent with the application, we focus on the estimands $\tau^{\TT}$ 
and
$\tau^{\PRTE}$, the finite sample version of which in the simulated dataset $b$ 
are calculated as follows,
\begin{eqnarray*}
	\tau^{\TT}_{b}&=& \sum_{i=1}^{N}  T_i \{Y_{i}(1)-Y_{i}(0)\}  \big/
	\sum_{i=1}^{N} T_{i},\\
	\tau^{\PRTE}_{b}&=&\sum_{i=1}^{N} \delta_{i} \{Y_{i}(1)-Y_{i}(0)\}\big/
	\sum_{i=1}^{N}{\delta_{i}}, \text{ with
	}\delta_{i}=\textbf{1}_{z_{\min}\leq
		S_{i}\leq z_{\max}}.
\end{eqnarray*}
We then apply the local IV method described in Section \ref{sec:Bayesian} and
OLS and 2SLS described in Section \ref{sec:compare}.

\begin{figure}[H]
	\centering
	\includegraphics[scale=0.45]{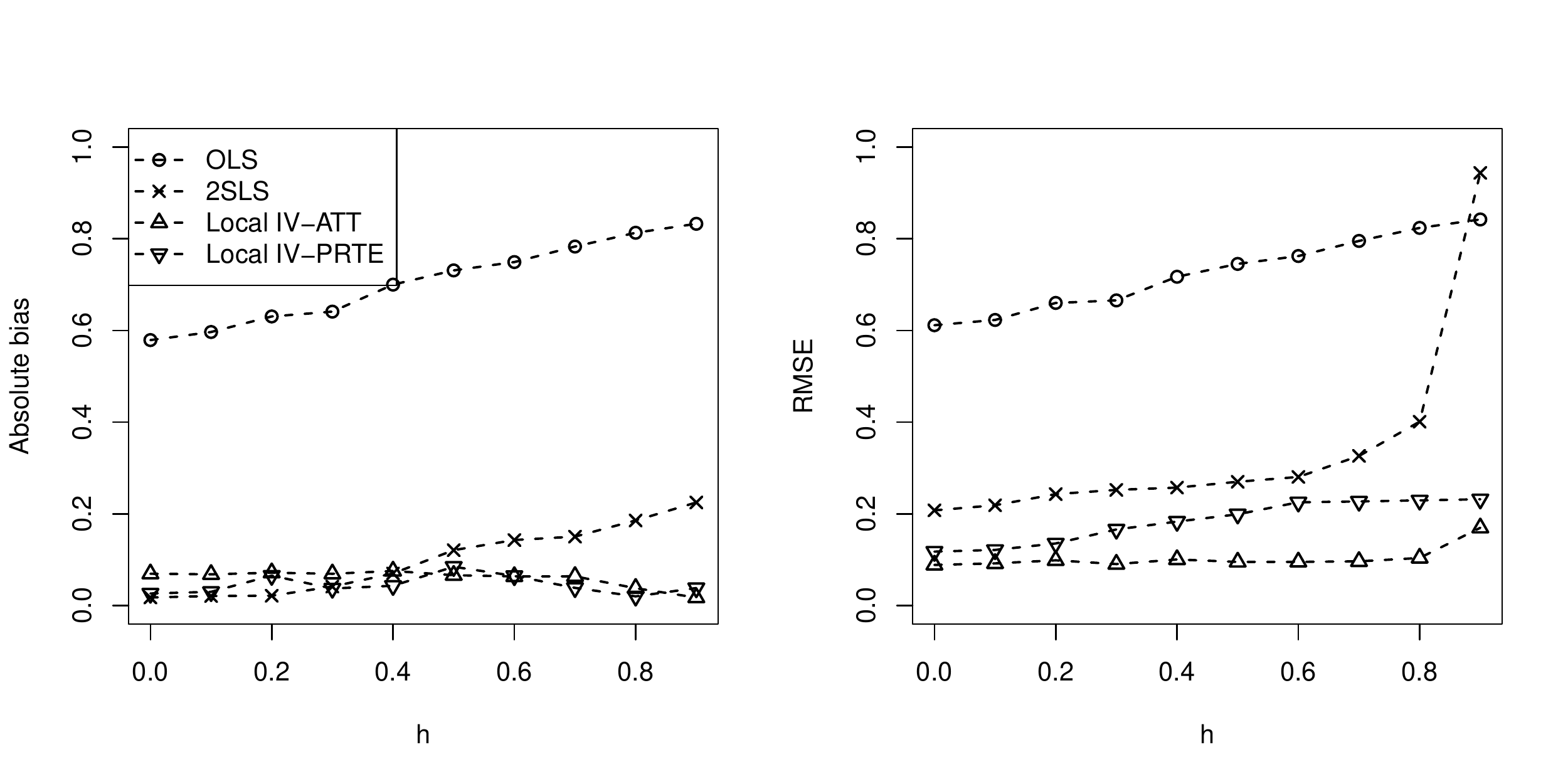}
	\caption{The absolute bias (left) and RMSE (right) of local IV (estimating 
		$\tau^{\TT}$ and
		$\tau^{\PRTE}$), OLS and 2SLS estimates under different degree of 
		treatment effect heterogeneity, $h$, in the simulation studies. }
	\label{fig:heter_simu}
\end{figure}

We fix the parameters $b_{00}=0,b_{01}=1,b_{10}=3,b_{11}=2,
p=0.5$ and sample size $N=1000$, and vary the degree of
heterogeneity $h$. For each setting, we simulate 1000 replicates and apply
local IV, OLS and 2SLS to each replicate. We calculate the absolute bias and
squared root of the mean squared error (RMSE) of $\tau^{\TT}$ and
$\tau^{\PRTE}$ of each method. Figure \ref{fig:heter_simu} displays the 
absolute bias and RMSE, respectively, between the local IV (for
$\tau^{\TT}$ and $\tau^{\PRTE}$), OLS, and 2SLS estimates under different
degree of heterogeneity $h$. We can see that the local IV approach obtains 
lower bias and
RMSE compared with OLS and 2SLS in estimating both $\tau^{\TT}$ and
$\tau^{\PRTE}$. Moreover, as the treatment effects become more heterogeneous 
between different principal strata,
the advantage of local IV over OLS and 2SLS in terms of bias and RMSE becomes 
larger.
This supports the use of the local IV method in settings with heterogeneous
treatment effects, which is likely to be the case in the OCP application.

\section{Conclusion} \label{sec:discuss}

Leveraging the different implementation intensity of the OCP as a natural 
experiment, we evaluated the causal effects of being an only child on 
self-reported psychological health measures using data from the China Family 
Panel Studies. We found small but significant negative effects of being an only 
child. We also found two sources of treatment effect heterogeneity. First, the 
negative effects are more pronounced among those from  urban areas. Second, the 
effects decrease with the latent degree of family resistance to the policy, 
characterized by a selection model of the family decision. Our results support 
the importance of sibship in psychological development. In the context of 
quantity-quality trade-off, our results suggest that the decreasing family size 
or the lack of sibship is not necessarily beneficial to individuals' 
psychological development. Namely, the ``quality'' of child, measured by the 
psychological health in our application, does not improve as the ``quantity'' 
decreases. Future work would investigate the possible trade-off between family 
size or the number of children and the well-being of children from other 
aspects.

From a methodological perspective, we made several extensions of the local IV 
method for continuous IV analysis \citep{Heckman1999}. We elucidate an 
intrinsic connection between local IV and principal stratification, which 
allows us to employ Bayesian hierarchical models to accommodate complex data 
structure such as clustering. Within the same framework, one could also adopt 
more flexible Bayesian semiparametric or nonparametric models. For simplicity, 
we focused on average causal estimands for ordinal outcomes, while 
acknowledging the literature on ordinal outcome in the presence of 
noncompliance \citep{Cheng2009,Baker2011} and 2SLS models for ordinal outcomes 
\citep{miranda2006maximum}. We have also explored nonadditive estimands as the 
ones proposed in \cite{Lu2015} and found similar patterns as the average 
effects.

\section*{Acknowledgements}
The authors are grateful to the editor of the special issue and two anonymous 
reviewers for constructive comments that help improve the clarity and 
exposition of the article. We also benefited from discussions with Liyun Chen, 
Chaoran Yu, Wentao Xiong, Zhichao Jiang, Wei Li, Yi Zhou, Wei Huang, and Yu Xie.
Peng Ding was supported by the U.S. National Science Foundation (grants \# 
1713152 and \# 1945136).

\bibliographystyle{jasa3}	 
\bibliography{Reference}

\end{document}